\documentclass[prc,twocolumn,preprintnumbers,showpacs]{revtex4}
\usepackage{amsmath,amssymb,epsfig,bm,mathrsfs,feynmp,color}
\usepackage{nicefrac}
\usepackage{exscale} 
\usepackage[colorlinks=true,linkcolor=blue,citecolor=cyan,urlcolor=blue]{hyperref}
\newcommand{\be}{\begin{equation}}
\newcommand{\ee}{\end{equation}}
\newcommand{\bea}{\begin{eqnarray}}
\newcommand{\eea}{\end{eqnarray}}

\def\apj{ApJ}%
\def\aap{A\&A}%
\def\jcap{J. Cosmology Astropart. Phys.}%
\def\mnras{MNRAS}%
\def\na{New A}%
\def\prc{Phys.~Rev.~C}%
\def\prd{Phys.~Rev.~D}%
\begin{document}

\title{Axion cooling of neutron stars} 
\begin{abstract}
  Cooling simulations of neutron stars and their comparison with the
  data from thermally emitting x-ray sources put constraints on the
  properties of axions, and by extension of any light pseudoscalar
  dark matter particles, whose existence has been postulated to solve
  the strong-CP problem of QCD. We incorporate the axion emission by
  pair-breaking and formation processes by $S$- and $P$-wave nucleonic
  condensates in a benchmark code for cooling simulations as well as
  provide fit formulas for the rates of these processes. Axion cooling
  of neutron stars has been simulated for 24 models covering the mass
  range 1 to 1.8 solar masses, featuring nonaccreted iron and
  accreted light-element envelopes, and a range of nucleon-axion
  couplings. The models are based on an equation state predicting
  conservative physics of superdense nuclear matter that does not
  allow for the onset of fast cooling processes induced by phase
  transitions to non-nucleonic forms of matter or high proton
  concentration. The cooling tracks in the temperature vs age plane
  were confronted with the (time-averaged) measured surface
  temperature of the central compact object in the Cas A supernova
  remnant as well as surface temperatures of three nearby middle-aged
  thermally emitting pulsars.We find that the axion coupling is
  limited to $f_a/10^{7}\textrm{GeV} \ge (5$--$10)$, which translates into
an  upper bound on axion mass $m_a \le (0.06$--$0.12)~\textrm{eV}$ for
  Peccei-Quinn charges of the neutron $\vert C_n \vert \sim 0.04$ and proton
 $\vert C_p \vert \sim 0.4$ characteristic for hadronic models of axions.
\end{abstract}

\author{ Armen Sedrakian}
\affiliation{Institute for Theoretical
   Physics, J.~W.~Goethe-University, D-60438 Frankfurt-Main, Germany}

\maketitle

\section{Introduction}
\label{sec:intro}

Astrophysics provides means for constraining properties of dark matter
particles -- in particular, light pseudoscalar particles such as {\it
  axions}~\cite{1978PhRvL..40..279W,1978PhRvL..40..223W}.  Axions were
originally introduced in the context of the Peccei-Quinn mechanism
which postulates a new global $U(1)_{PQ}$ symmetry
~\cite{1977PhRvL..38.1440P,2008LNP...741....3P} to solve the strong-CP
problem in QCD~\cite{1976PhRvL..37....8T}, but they may play a
significant role in cosmology and in stellar physics.  Stellar physics
of the Sun and solar type stars, red giants, white dwarfs and
supernovae puts constraints on the couplings of axions to
standard-model (SM) particles~\cite{2011PhRvD..84j3008R}. The
constraints are set by requiring that the coupling of axions to SM
particles not alter significantly the agreement between theoretical
models and observations.  Axions may efficiently be produced in the
interiors of stars and act as an additional sink of energy; therefore,
they can alter the energetics of some processes -- for example, a
type-II supernova explosion.  Several authors noted that the emission
of axions ($a$) in the nucleon ($N$) bremsstrahlung $N+N\to N+N+a$ may
drain too much energy from the type-II supernova process, making it
energetically inconsistent with observations of such
events~\cite{1988PhRvD..38.2338B,1989PhRvD..39.1020B,1990PhRvD..42.3297B,1996PhRvL..76.2621J,2001PhLB..499....9H}. Axions
will not affect the neutrino burst if they are trapped inside the
newborn neutron star, which would be the case if the axion mass is
larger than $10^{−2}$~eV~\cite{1990PhRvD..42.3297B}. In this case the
axions are radiated, in analogy to neutrinos, from the ``axion
sphere.''  Combined studies of the free-streaming and trapping regimes
suggest that an axion with mass in the interval $10^{−3}$ to 2~eV is
excluded by the observation of neutrinos from SN
1987A~\cite{1990PhRvD..42.3297B}. The coupling of axions to other SM
particles is also constrained by stellar physics. For example, the
axion coupling to electrons is constrained by the cooling of white
dwarfs and red giants, where the underlying energy-loss mechanism is
the axion emission by bremsstrahlung of electrons scattering off
nuclei~\cite{1994APh.....2..175A,1995PhRvD..51.1495R,2001NewA....6..197C,2014JCAP...10..069M}. Solar
physics provides another example where energy arguments allow us to
place limits on beyond-SM physics; see
Refs.~\cite{2015JCAP...10..015V,2013JCAP...12..008R}.  These stellar
constraints are complemented by
experimental~\cite{2013JCAP...05..010B} and
cosmological~\cite{2013JCAP...10..020A} bounds.  For reviews of
astrophysical limits on axion properties, see
Refs.~\cite{2008LNP...741...51R,2010RvMP...82..557K}.

Neutron star cooling by neutrino emission is a highly sensitive tool
to study the interior composition of neutron stars~(see, for example,
reviews~\cite{weber_book,2007PrPNP..58..168S,2013arXiv1302.6626P}).
Neutron star cooling via axions has evaded detailed scrutiny, although a
number of key reactions necessary for such an analysis have been
computed long
ago~\cite{1984PhRvL..53.1198I,2001PhRvD..64d3002I,1987ApJ...322..291N,1988ApJ...326..241N}
(for details, see Sec.~\ref{sec:emission_processes}). Umeda
{\it et~al.} \cite{1998nspt.conf..213U}, in their pioneering study of axion
cooling of neutron stars, considered the axion radiation process via the
bremsstrahlung in $NN$ collisions in bulk nuclear matter. However,
neutrino-antineutrino pair emission via Cooper
pair-breaking-formation (PBF)
processes~\cite{1976ApJ...205..541F,1987PhLB..184..119S}, which start
to operate below the critical temperature of transition of baryons to
the superfluid state, plays an important role in the modern simulations
of cooling of neutron stars.  These processes act as the dominant cooling agent
during the {\it neutrino cooling era} (i.e., the time span
$0.1\le t\le 100$~kyr) if the fast cooling processes are not operative.
Previously, Ref. \cite{2013NuPhA.897...62K} (hereafter abbreviated as
KS) computed the axion counterparts of the PBF processes in neutron
stars and set approximate limits on the axion's coupling to baryons and
its mass by requiring that the axion emission rate via the PBF
processes be smaller than its neutrino
counterpart~\cite{1999A&A...343..650Y,2006PhLB..638..114L,2007PhRvC..76e5805S,2008PhRvC..77f5808K,2012PhRvC..86b5803S}.

The purpose of this work is to continue the KS analysis by
incorporating the rates of the PBF processes in a cooling simulation
code. Here we compute a large sample of cooling models of neutron
stars and confront them with observations.  The first aspect of our
strategy is to use a {\it conservative model} of cooling which is not
contaminated by the uncertainties in the rates of rapid neutrino
emission processes, which in turn strongly depend on the composition
of dense matter at densities above the saturation of nuclear
matter. Modern simulations of cooling of neutron stars (see, for
example, the work by different groups on hadronic models
\cite{2009ApJ...707.1131P,2011MNRAS.412L.108S,2012PhRvC..85b2802B,2014JPhCS.496a2014G,2013MNRAS.434..123V,2015MNRAS.446.3621S}
and hybrid star models~\cite{2011PhRvD..84f3015H,
  2013A&A...555L..10S,2015arXiv150906986S, 2015PhRvC..92c5810D})
demonstrate that fast neutrino processes do not operate in low-mass
neutron stars with $M\le 1.5 M_{\odot}$ because each such process is
associated with a certain density threshold (which need not be sharp,
see in particular Refs.~\cite{2012PhRvC..85b2802B,2014JPhCS.496a2014G}
for this type of modeling). Light neutron stars may not achieve these
thresholds in their centers, and therefore they will follow the {\it
  slow cooling} scenario which is in line with the {\it minimal
  cooling paradigm} that excludes fast cooling processes {\it per
  se}~\cite{2009ApJ...707.1131P}.  Below, the axion bounds will be
derived from simulations of the cooling of low-mass stars.  (We will also
report results obtained for more massive stars in the framework of
minimal cooling, i.e., by simply excluding the fast processes, such
as the direct Urca process.) The Akmal-Pandharipande-Ravenhall (APR) 
equation of state (EOS) that will be used in our
simulations has nucleons and leptons as constituents of matter at all
densities and does not include non-nucleonic degrees of 
freedom~\cite{2009ApJ...707.1131P}.

The second aspect of our strategy is to concentrate on a small sample
of relatively high-temperature young and intermediate-aged objects
which reside within the time domain $0.1\le t\le 100$ kyr and which
are known to be weakly magnetized.  The latter choice guarantees that
no contamination will arise from the uncertain physics of internal
heating processes. As argued in KS, a single example that does not fit
into the axion cooling scenario already constrains the coupling of axions
to SM particles.  As a representative for young nonmagnetized neutron
stars we choose the compact central object (CCO) located in the Cas A
supernova remnant. As with all CCOs, this neutron star emits radiation in
x rays without counterparts at other wavelengths. As a representative
for intermediate-aged neutron stars, we selected three nearby thermally
emitting neutron stars, two of which are radio-active pulsars B0656+14
and B1055-52, with the third being the radio-quiet neutron star Geminga.

Finally, we use a benchmark
code~\cite{Comment} which incorporates
standard microphysical input (EOS, gaps, etc.) used commonly in the
cooling simulations. For details of the code, physics input and
results, see Ref.~\cite{2009ApJ...707.1131P} and references
therein. We also conducted simulations with an alternative code
described in
Refs.~\cite{2011PhRvD..84f3015H,2013A&A...555L..10S,2015arXiv150906986S},
with different EOSs and microphysics input, and obtained
quantitatively good agreement at the relevant intermediate- and
late-time cooling.

This paper is structured as follows: In Sec.~\ref{sec:axions} we
review the axion properties and their emission rates in neutron stars.
The cooling simulations and the results are discussed in
Sec.~\ref{sec:cooling}.  Our conclusions and an outlook are given in
Sec.~\ref{sec:conclusions}.

\section{Axion emission rates in neutron stars}
\label{sec:axions}

\subsection{Axion couplings to SM particles}
\label{sec:axion_SM}

Quantum chromodynamics (QCD), the fundamental theory of strong
interactions, violates the combined CP symmetry due to a topological
interaction term in the QCD Lagrangian
\be
\label{thetaaction}
\mathscr{L}_\theta={g^2\theta\over 32\pi^2}
 \,F^a_{\mu\nu}\tilde F^{\mu\nu a}, 
\ee
where
$F_{\mu\nu}^a = \partial_{\mu}A_{\nu}-\partial_{\nu}A_{\mu}
+gf^{abc}A_{\mu b} A_{\nu c}$
is the gluon field strength tensor, $g$ is the strong coupling
constant,
$\tilde F^a_{\mu\nu} = \epsilon_{\mu\nu\lambda\rho} F^{\lambda\rho
  a}/2$,
$f^{abc}$ are the structure constants of the $SU(3)$ group, and the parameter
$\theta$, which is periodic with period $2\pi$,  parametrizes the
nonperturbative vacuum states of QCD
$\vert\theta\rangle = \sum_n \exp(-in\theta)\vert n\rangle$; here $n$
is the winding number characterizing each distinct state, which is not
connected to another by any gauge transformation~\cite{1976PhRvL..37....8T}.
If quarks are present, then the physical parameter is
$
\bar\theta = \theta + \arg\det m_q , $
 where $m_q$ is the matrix of
quark masses. Experimentally, the upper bound on the value of this
parameter is $\bar\theta\lesssim 10^{-10}$, which is based on the
measurements of the electric dipole moment of
the neutron,  $d_n<6.3\cdot
10^{-26}e$ cm~\cite{2006PhRvL..97m1801B}.  SM does not provide an
explanation on why $\bar\theta$ is not of the order of unity -- a fact
known as the strong CP problem.

The Peccei-Quinn mechanism solves the CP problem by introducing an new
global $U(1)_{PQ}$ symmetry which adds an additional anomaly term to
the QCD action proportional to the axion field $a$
~\cite{1977PhRvL..38.1440P,1978PhRvL..40..279W}.  The axion
field value is then given by 
$\langle a \rangle \sim -\bar\theta$.  The physical axion field is
then $a- \langle a \rangle $, and the undesirable $\theta$ term in
the action is replaced by the physical axion field, which can be
viewed as the
Nambu-Goldstone boson of the Peccei-Quinn $U(1)_{PQ}$ symmetry
breaking~\cite{1978PhRvL..40..223W,1978PhRvL..40..279W}.

The Lagrangian of axion field $a$ has the form
\be \mathscr{L}_a = -\frac{1}{2}\partial_{\mu}
a\partial^{\mu} a + \mathscr{L}^{(N)}_{int}(\partial_{\mu} a,\psi_{N})
+ \mathscr{L}^{(L)}_{int}(a,\psi_L),
\ee
where the second term describes the coupling of the axion to nucleon
fields ($\psi_N$) and the third term describes the coupling to the lepton
fields ($\psi_L$) of the SM.  The coupling of axion to nucleonic fields
is described by the following interaction Lagrangian:
\be 
\mathscr{L }^{(B)}_{int} = \frac{1}{f_a} B^{\mu}  A_{\mu},
\ee
where $f_a$ is the axion decay constant, and the baryon and axion currents 
are given by
\be\label{eq:currents}
B^{\mu} =
\sum_{N} \frac{C_N}{2}  \bar\psi_N\gamma^{\mu}\gamma_5\psi_N,\quad \quad 
A_{\mu} = \partial_{\mu} a,
\ee
where $N\in n,p$ labels neutrons and protons, and $C_N$ are the
Peccei-Quinn (PQ) charges of the baryonic currents. The dimensionless
Yukawa coupling can be defined as $g_{aNN} = C_Nm_N/f_a$ with 
the implied 
``fine-structure'' constant $\alpha _{aNN} = g^2_{aNN}/4\pi$. The
coupling of axions to leptons (in practice we consider only electrons)
is commonly taken in the pseudoscalar form 
\bea
\mathscr{L}^{(e)}_{int}(a,\psi_e) = -i g_{aee} \bar
\psi_e\gamma_5\psi_e a, 
\eea 
where the Yukawa coupling is given by
$g_{aee}= C_em_e/f_a$.  The $C_N$ charges are generally given by
generalized Goldberger-Treiman relations 

\bea C_p & =& (C_u-\eta)
\Delta_u+(C_d-\eta z) \Delta_d +(C_s-\eta w)\Delta_s,\\
C_n & =& (C_u-\eta) \Delta_d+(C_d-\eta z) \Delta_u 
+(C_s-\eta w)\Delta_s, 
\eea 
where $\eta = (1+z+w)^{-1}$, with $z = m_u/m_d$, $w = m_u/m_s$, and
$\Delta_u =0.84\pm 0.02$, $\Delta_d = -0.43\pm 0.02$ and
$\Delta_s = -0.09\pm 0.02$. The main uncertainty is associated with
$z = m_u/m_d = 0.35$--$0.6$. For {\it hadronic axions}, $C_{u,d,s} = 0$,
and the nucleonic charges vary in the range
\bea \label{eq:axion_range} -0.51 \le C_p\le -0.36, \quad -0.05 \le
C_n\le 0.1 .  
\eea 
These ranges imply that neutrons may not couple to
axions ($C_n=0$) whereas protons always couple to axions $C_p\neq 0$.
The values of PQ charges define a continuum of axion models; for a
review see, for example, Ref.~\cite{2008LNP...741....3P}.  In the so-called
{\it invisible axion} DFSZ model, these couplings are of the same order
of magnitude and are related via the ratio of two Higgs vacuum
expectation values $\tan\beta$ as follows:
$ C_e = {\cos^2\beta}/{3},\quad C_u = {\sin^2\beta}/{3}, \quad C_d =
{\cos^2\beta}/{3}, $
where $\beta$ is a free parameter. In the alternative KVSZ model
ordinary, SM particles do not have PQ charges and $C_e = 0$; the
coupling of baryons to axions arises from PQ charges of unknown very
heavy quarks. To keep the discussion general enough, we will abstract
from a particular axion model and will treat the PQ charges of
fermions as free parameters taken from the range
\eqref{eq:axion_range}; we will also explore the case of large neutron
PQ charge to contrast our result with the case where
$\vert C_n\vert \sim \vert C_p\vert$.  If only nucleonic processes are
considered, the emission rates depend on a certain combination of
charges and axion decay constant. In general, when leptonic processes 
are involved, this is not the case.

The axion mass is  related to $f_a$ via the relation 
\bea 
\label{eq:axion_mass}
 m_a = \frac{z^{1/2}}{1+z} \frac{f_\pi m_\pi}{f_a} = \frac{0.6 
  ~\textrm{eV}}{f_a/10^{7}~\textrm{GeV}} 
\eea 
where the pion mass $m_\pi=135$ MeV, decay constant $f_\pi = 92$ MeV,
and we adopt from the range of $z$ values quoted  the 
value $z = 0.56$. Equation~(\ref{eq:axion_mass}) translates a lower
bound on $f_a$  into an upper bound on the axion mass.

\subsection{Axion emission via PBF process}
\label{sec:emission_processes}

KS obtained the axion emissivity of $S$-wave paired superfluid by
assuming that the PQ charges of nucleons are fixed by
 $C_N/2 = 1$. By matching 
Eq. (5) of KS with Eq.~\eqref{eq:currents} we see that we need to
rescale their $f_a^{-1} \to (C_N/2)f_a^{-1}$ to obtain explicitly the
expression for the axion emissivity in the present notations. Thus,
the axion emissivity now reads 
\bea\label{eq:axion1} \epsilon^S_{aN}  &=& \frac{ 2 C^2_N}{3\pi} \, f_a^{-2}
\,\nu_N(0)\, v_{FN}^2 \, T^5 \, I^S_{aN}, 
\eea 
where  $\nu_N(0) = m_N^*p_{FN}/\pi^2$  is the density of states at the Fermi
surface, $v_{FN}$ is the Fermi velocity, the superscript $S$ indicates
isotropic pairing in the $^1S_0$ channel,
\bea \label{eq:Ia}
I^S_{aN} =
z_N^5\int_1^{\infty}\!\! dy ~ \frac{y^3}{\sqrt{y^2-1}} f_F\left(z_N 
  y\right)^2, 
\eea 
and $z_N= \Delta^S_N(T)/T$. The bound obtained by KS from the
requirement that the axion cooling not overshadows the cooling via
neutrinos after rescaling reads
\bea \label{eq:ratio2}
\frac{\epsilon^S_a}{\epsilon^S_{\nu}} &=& 
\frac{59.2C_N^2}{4 f_a^{2} G_F^2 \Delta^S_N(T)^2}~r(z) \le 1
\eea
where $r(z)$ is the ratio of the phase-space integral for axions
(\ref{eq:Ia}) and its counterpart for neutrinos and is numerically
bound from above $r(z)\le 1$; therefore it can be dropped from the
bound on $f_a$.  Substituting the value of the Fermi coupling
constant $ G_F = 1.166\times 10^{-5}$ GeV$^{-2}$ in
Eq.~(\ref{eq:ratio2}),  we rewrite the bound found by KS as
\be\label{eq:fbound} 
\frac{f_a/10^{10}\textrm{GeV}}{C_N} >  0.038 
\left[\frac {1~\textrm{MeV}}{\Delta^S(T)}\right].
\ee
which now includes the PQ charge of the neutron or proton explicitly.
Using Eq.~(\ref{eq:axion_mass}), this  translates to an upper bound
on the axion mass of
\be \label{eq:mbound} 
m_a \, C_N\le 0.163 ~  \textrm{eV}\,\left(\frac {\Delta^S_N(T)}{1~\textrm{MeV}}\right).
\ee 
Note that the nucleon pairing gap on the right-hand side can be
replaced by the critical temperature $T_c$, because in the range of
temperatures important for pair-breaking processes, $0.5\le T/T_c< 1$
the BCS theory predicts $\Delta (T) \simeq T_c$.

Neutron condensate in neutron star cores is paired in the
$^3P_2$-$^3F_2$ channel in a state which features an anisotropic
gap~\cite{2003NuPhA.720...20Z}. As pointed out in KS, the results above
can be trivially extended to the $P$-wave pairing following analogous
discussion for neutrino emission in Ref.~\cite{1999A&A...343..650Y}.
The corresponding axion emissivity is obtained from the $S$-wave rate
above (\ref{eq:axion1}) by setting $v^2_{Fn} = 1$ and angle-averaging
the phase-space integral (\ref{eq:Ia}) to account for the anisotropy of
the gap
\bea\label{eq:axion2}
\epsilon^P_{an} &=& \frac{ 2 C^2_n}{3\pi} \, f_a^{-2} \,\nu_n(0)\,
\, T^5 \, I^P_{an}, 
\eea
where 
\bea 
\label{eq:IaP} I^P_{an} =\int \frac{d\Omega}{4\pi}
z_N^5\int_1^{\infty}\!\! dy ~ \frac{y^3}{\sqrt{y^2-1}} f_F\left(z_N 
  y\right)^2, 
\eea 
where $d\Omega$ denotes the integration over the solid angle and $z_N
= \Delta^P(T, \theta)/T$ depends on the polar angle $\theta$, where
$\Delta^P(T, \theta)$ is the pairing gap in the $P$-wave channel.
By adapting Eq.~(\ref{eq:Ia}) to the $P$-wave case, we automatically include 
the vertex corrections that were omitted in
Ref.\,\cite{1999A&A...343..650Y}.  Note that $C_n=0$ is not excluded;
i.e., conceivably axions may not be emitted by the neutron $P$-wave
condensate.

For the purpose of numerical simulations of axion cooling, it is useful
to obtain fits to the dependence of the integrals (\ref{eq:Ia}) and
(\ref{eq:IaP}) on reduced temperature $\tau = T/T_c$, where $T_c$ is
the critical temperature. We first obtain the asymptotic forms of
these integrals in the limits $T\to 0$ and $T\to T_c$.  In the
low-temperature limit $\Delta(T)/T \gg 1$, i.e., $z\gg 1$, and because
$y\ge 1$ we can set in (\ref{eq:Ia})
$f^2_F\left(z y\right) = \exp{(-2yz)}$. (We drop the indices $N,n$ and
$p$ in the intermediate steps and recover them in the final
expressions).  The integration with subsequent expansion in $z\gg 1$
gives
\bea \label{eq:Ia_lowT}
I^S_{aN} &= &
z^5\left[ K_1(2z) +\frac{ K_2(2z)}{2z}\right]
\simeq \frac{z^5}{2} \sqrt{\frac{\pi}{z}} \exp (-2z),\nonumber\\
\eea
where $K_n(z)$ is the Bessel function of the second kind of $n$th order.

In the limit $T\to T_c$, we approximate the denominator of the
integrand $\sqrt{y^2-1}\simeq y$ and obtain
\bea 
\label{eq:Ia_highT}
I^S_{aN}&=&  \left(9\zeta(3)-\pi^2\right)\frac{\Delta_N^2}{6T^2} \simeq
0.158 \frac{\Delta_N^2}{T^2}.
\eea
There exist two competing states for $P$-wave superfluid, which differ
by the anisotropy of the gap.  We denote these states as A and B and assign
them the following dependences on the angle $\theta$:
\bea 
\Delta^A = \Delta_0^A\sqrt{1+3\cos^2\theta}, 
\quad 
\Delta^B = \Delta_0^B\sin\theta. 
\eea 
In the high-temperature ($T\to T_c$) limit we have 
\bea 
I_{an}^{P ^ A} &=& I^{P}_{an}\int \frac{d\Omega}{4\pi}
(1+3\cos^2\theta)= 2 I^{P^0}_{an},\\
I_{an}^{P ^ B} &=& I^{P}_{an}\int \frac{d\Omega}{4\pi} \sin^2\theta
= \frac{2}{3}  I^{P^0}_{an} ,
\eea 
where $I^{P^0}_{an}$ stands for the isotropic part of the integral and
is given by (\ref{eq:Ia_highT}) where $\Delta_N$ is replaced by
$\Delta_0^{A,B}$.  In the anisotropic case, the low-temperature limit
does not have a simple analytical representation.

\begin{figure}[t] 
\begin{center}
\includegraphics[width=8cm,height=7cm]{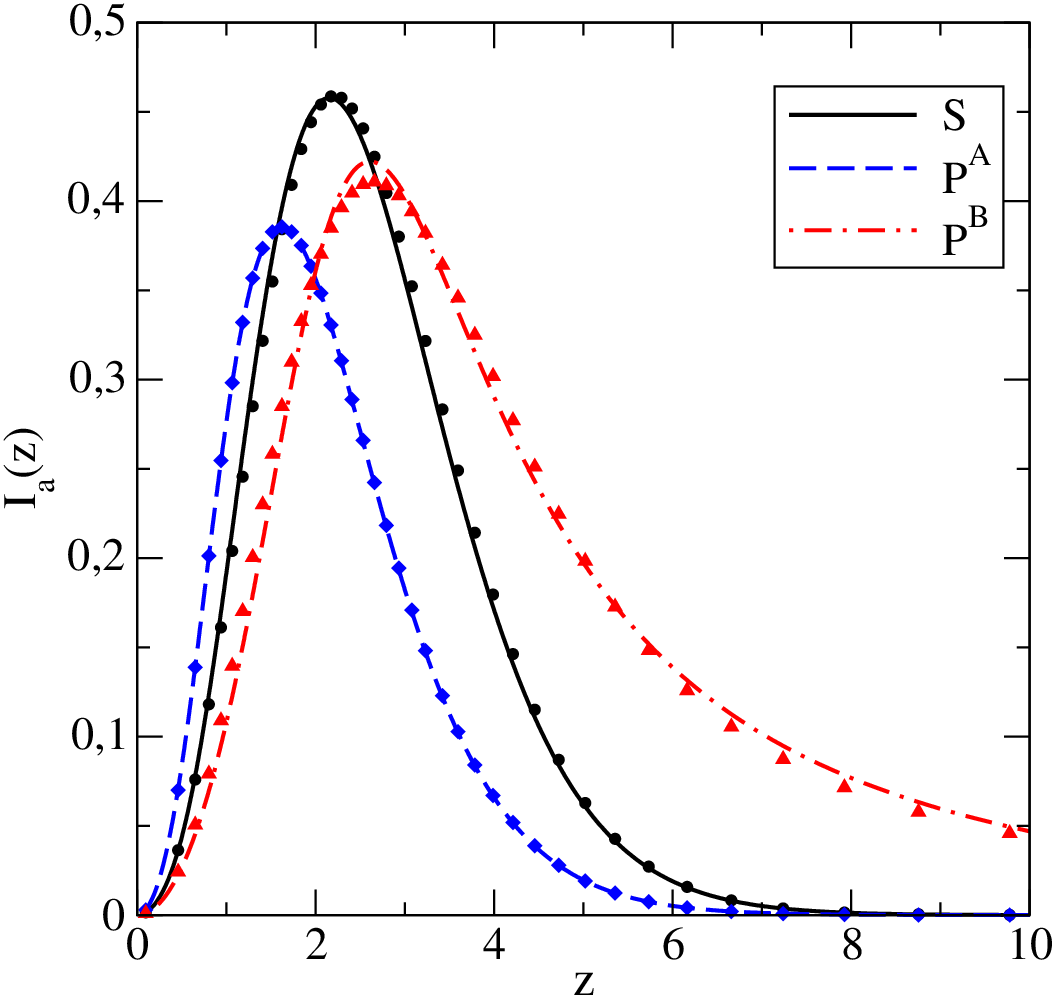}
\caption{   Dependence of the integrals \eqref{eq:Ia} and \eqref{eq:IaP}
on   $z = \Delta(T)/T$. The exact results are shown by symbols, whereas the 
  fits are shown by lines.  }
\label{fig:1}
\end{center}
\end{figure}

The exact numerical calculations of the integrals (\ref{eq:Ia})
and (\ref{eq:IaP}) were fitted in the range $0\le z\le 15$ using suitable functions which
reproduce  correct asymptotic forms as described above.
For $S$-wave pairing, we used the following fit formula:
\bea \label{eq:fit_IS}
I^S_{aN}(z) &=&  (az^2 + cz^4) \sqrt{1 + fz}\,
e^{-\sqrt{4 z^2 + h^2} + h},
\eea
where $a = 0.158151$, $c = 0.543166$, $h = 0.0535359$, and
$f = \pi/4c^2$.  This formula fits the numerical result with relative
accuracy $\le 5.6 \%$ for $z \sim 1$ and much more accurately in the
asymptotic regimes.  In the case of $P^A$-wave pairing, we used the
function
\bea \label{eq:fit_IPA} I^{P^A}_{an} (z) &=& \frac{(az^2 + cz^4) (1 +
  fz^2)}{(1+bz^2+gz^4)} e^{-\sqrt{4 z^2 + h^2} + h}, \eea where
$a = 2\times 0.158151$, $b = 0.856577$, $c = 0.0255728$,
$f = 2.22858$, $g = 0.000449543$ and $h = 2.22569$. The relative
accuracy of the fit is $\le 4\%$ at $z\sim 10$ and is better in the
rest of the domain.  Finally, in the case of $P^B$-wave pairing we
used the function
\bea \label{eq:fit_IPB}
I^{P^B}_{an}(z) &=&  \frac{(az^2 + cz^4) \sqrt{1 + fz^2}}{(1+bz^2+gz^4)} ,
\eea
where $a = (2/3)\times 0.158151$, $b = -0.043745$, 
 $c = -0.000271463$,  $f = 0.0063470221$,  $g = 0.0216661$. The relative
 error in this case remains below $2\%$. The exact
 results for the integrals \eqref{eq:Ia} and \eqref{eq:IaP} are shown
 in Fig.~\ref{fig:1} together with the approximate fits given by
Eqs.~\eqref{eq:fit_IS}--\eqref{eq:fit_IPB}.

\subsection{Axion bremsstrahlung emission in the crust}

Electrons undergoing acceleration in the vicinity of a nucleus characterized by
charge $Z$ and mass number $A$ will emit axions. The 
PQ charge of electrons $C_e$ is related to the  dimensionless coupling 
of axions to electrons by $g_{aee} = {C_em_e}/{f_a}$.
 The emissivity of the axion bremsstrahlung 
process is given by~\cite{1984PhRvL..53.1198I,1987ApJ...322..291N}
\bea \label{eq:aee}
\epsilon_{aee}= \frac{\pi^2}{120} 
\frac{Z^2\alpha}{A } \left(\frac{C_em_e}{f_a\epsilon_e}\right)^2 n_B T^4 
\left[2   \ln (2\gamma) -\ln \frac{\alpha}{\pi}\right],
\eea  
where $\epsilon_e$ is the Fermi energy of electrons and $\gamma$ is
the Lorentz factor of ultrarelativistic electrons, $\alpha = 1/137$
is the fine-structure constant, and $n_B$ is the baryon number
density. This axion bremsstrahlung process has its neutrino-pair
emission counterpart and its rate is given
by~\cite{1999A&A...343.1009K}
\bea 
\epsilon_{\nu ee} =\frac{8\pi}{567} G_F^2 C_{+}^2 Z^2\alpha^4 n_i T^6 L ,
\eea 
where $C_+^2 = 1.675$, $0\le L\le 1$ includes many-body corrections
to the rate of the process related to the correlations among the
nuclei, electron screening, finite nuclear size, etc., and $n_i$ is the
number density of nuclei.  To see the relative importance of the axion
and neutron emissivities, we fix the electron PQ charge $C_e=1$, in
which case the ratio of the axion to neutrino emissivity is given by
\bea
\frac{\epsilon_{a ee}}{\epsilon_{\nu ee}}  
&\simeq& \frac{189\pi}{320 }\frac{C_e^2}{(C_+G_FTf_a)^2\alpha^3 L}
\left(\frac{m_e}{\epsilon_e}\right)^2
\nonumber\\
& =  & 2.8
\left(\frac{1}{T/10^9\textrm{K}}\right)^2
\left(\frac{1}{f_a/10^{10}\textrm{GeV}}\right)^2,
\eea
where for the sake of estimate we set $m_e/\epsilon_e = 10^{-2}$,
$L=1$ and set the expression in brackets in Eq.~\eqref{eq:aee} equal to
unity. We also use $n_B/A = n_i$, which applies when the density of
free neutrons in the crust is negligible, as has been assumed in
Eq. (8) of Ref.~\cite{1984PhRvL..53.1198I}. 

\subsection{Axion bremsstrahlung emission in the core}

To describe the axion emission in the core of the neutron stars, we consider
the processes involving neutrons, protons and electrons; the EOS chosen
for numerical simulations is purely nucleonic for all relevant
densities, and there is no need to consider other degrees of freedom,
such as hyperons or quarks. Axions will be emitted in the nucleon
collisions via bremsstrahlung process (irrespective of the pairing of
nucleons). The emissivity of the process $N+N\to N+N + a$, $N\in n$ or
$p$, is given by ~\cite{1984PhRvL..53.1198I,2001PhRvD..64d3002I}
\bea
\label{eq:NNBrems}
\epsilon_{aN} = \frac{31}{945} \alpha_{aNN}
\left(\frac{f_\pi}{m_\pi}\right)^4 m_N^2p_{FN}\, T^6\,
F\left(\frac{m_\pi}{2p_{FN}}\right) {\cal R}, 
\eea 
where $\alpha_{aN}$ is the axion fine-structure constant (see Sec.
\ref{sec:axion_SM}), $F(x) \equiv 1-(3/2)x\arctan(1/x)+x^2/2(1+x^2)$.
We do not reproduce the expression for the $n+p\to n+p + a$ reaction, which
is more complicated due to two different Fermi surfaces involved, see
Eq.~(2.13) of Ref.~\cite{2001PhRvD..64d3002I}.  The factor ${\cal R}$
stands for reduction of the axion emissivity by the superfluidity of
nucleons and we have implemented the same factors as has been done for 
the neutrino bremsstrahlung in the code. (For discussion,
see Ref.~\cite{1995A&A...297..717Y}.) The emissivity \eqref{eq:NNBrems}
should be viewed as an upper limit because of the approximate treatment
of the nuclear interaction in the $N$-$N$ collisions which only
include the one-pion exchange contribution to the nuclear scattering
[hence the proportionality of $\epsilon_{aNN}$ to
$(f_\pi/m_\pi)^4$]. The inclusion of other (repulsive) channels of
interaction reduces the rate by a factor of 0.2. This argument applies
also to the neutrino-pair bremsstrahlung process in nuclear
collisions; therefore the relative importance of these processes in
cooling neutron stars is unaffected (i.e., the ratio of the axion and
neutrino emissivities is independent of the nuclear matrix element,
which can be factorized if the radiation is soft).

\section{Cooling simulations}
\label{sec:cooling}

We recall that the specific purpose of this work is to (a) consider a
conservative model of neutron stars without fast cooling agents which
is almost certainly guaranteed for light- to medium-mass neutron stars; (b)
choose observational data which are not potentially contaminated by the
heating by strong magnetic fields at intermediate stages of cooling
and other sources at late times; and (c) use a well-tested code with
standard EOS input in order to benchmark the axion cooling of neutron
stars and render the results easily reproducible.

\begin{figure}[tbh] 
\begin{center}
\includegraphics[width=8.4cm]{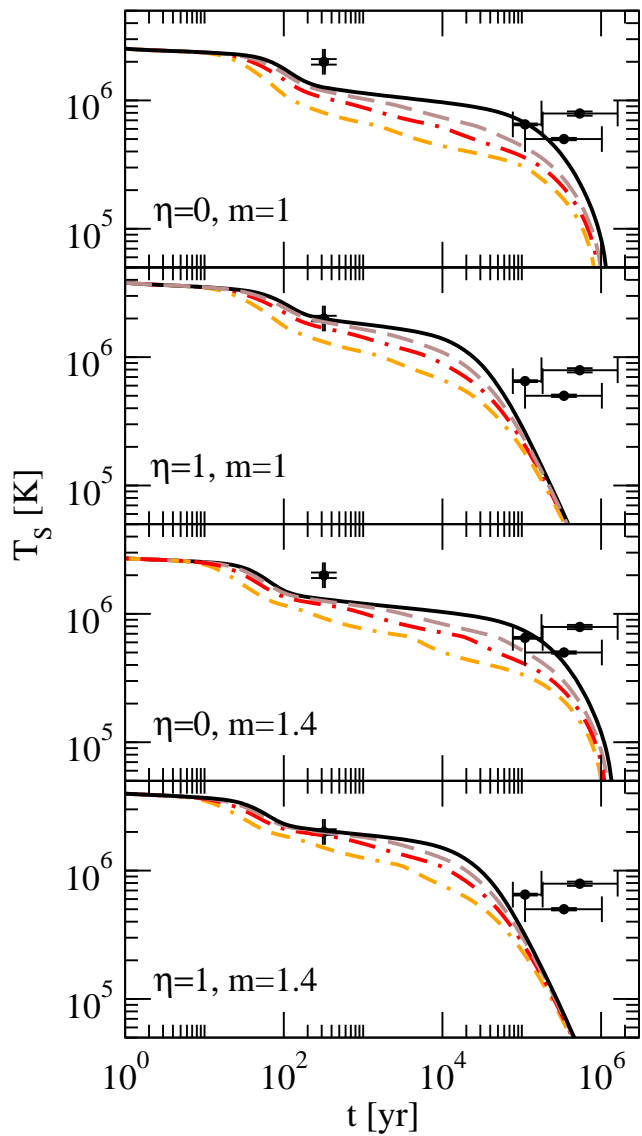}
\caption{   Cooling tracks ({\it redshifted} surface temperature vs age) 
  for neutron star models with masses $m=1$ and $m=1.4$ (in solar 
  units) for the cases of nonaccreted iron envelope ($\eta=0$) and 
  accreted light-element envelope ($\eta=1$). The representative 
  observational data includes (from left to right) the CCO in Cas A,
  PSR B0656+14, Geminga, and PSR B1055-52.  Each panel contains 
  cooling tracks for various values of the axion coupling constant;
  the case $f_a= \infty$ (solid line) corresponds to vanishing axion 
  coupling -- i.e., purely neutrino cooling. The axion cooling models 
  are shown for the values $f_{a7}= 10$ (dashed), $f_{a7}= 5$
  (dash-dotted), and $f_{a7}= 2$ (double-dash-dotted). }
\label{fig:2}
\end{center}
\end{figure}
\subsection{Physics input}

The cooling code solves the energy balance and transport equation, which
can be reduced to a parabolic differential equation for the
temperature of the core. The transport in the low-density blanket of
the star comprising matter below the density $\rho_b= 10^{10}$g
cm$^{-3}$ is decoupled from evolution and is treated separately in
terms of a relation between temperature at its base $T$ and the
surface of the star $T_s$. This relation has the generic form
$ T_s^4 = g_sh(T)$, where $g_s$ is the surface gravity, and $h$ is some
function which depends on $T$, the opacity of crustal material, and
its equation of state. The amount of the light material in the
envelope is regulated by the parameter $\eta$, which takes on the value
$\eta = 0$ for a purely iron surface and $\eta\to 1$ for  a light-element
surface. For a detailed discussion of the input physics, the reader is
referred to Ref.~\cite{2009ApJ...707.1131P} and references therein. 

After the initial nonisothermal phase the models settle into an
equilibrium state which is characterized by an isothermal core and
gradient-featuring envelope. In this case, the time evolution is
characterized by the ordinary differential equation
\be 
\label{eq:master} C_V \frac{dT}{dt} = -L_{\nu} (T)-L_{a} (T)
-L_{\gamma}(T_s)
+ H (T), 
\ee
where $L_{\nu}$ and $L_{a}$ are the neutrino and axion luminosities
from the bulk of the star (recall that the neutrino and axion
mean free paths are larger than the star radius) and $L_{\gamma}$ is
the luminosity of photons radiated from the star's surface.  Here $C_V$ is
the specific heat of the core, and $H(t)$ accounts for heating
processes, which could be important in the late-time evolution of
neutron stars. We assume below that $H(T) = 0$ in the neutrino cooling
era.  The photon luminosity is given simply by the Stefan-Boltzmann
law $L_{\gamma} = 4\pi \sigma R^2T_s^4$, where $\sigma$ is the
Stefan-Boltzmann constant, and $R$ is the radius of the star.

We have computed 24 models of cooling neutron stars by choosing three
different masses $m = 1.0$, 1.4, 1.8, where $m$ is the object mass
normalized to the solar mass, light-element $\eta= 1$ and iron
$\eta = 0$ envelopes, as well as four values of the axion decay
constant $f_{a7} = \infty, \, 10, \, 5, \, 2$, where we use the units
of $f_{a7} = f_a/10^7 $ GeV. Throughout most of the computation, the
PQ charges of neutrons and protons were fixed at
$\vert C_n \vert = 0.04$ and $\vert C_p\vert = 0.4$, which reflect the
asymmetry in the couplings of neutrons and protons to axions according
to Eq.~\eqref{eq:axion_range}. Note that these quantities enter the
axion emission rate in the combination
$(f^*_a)^{-1} = (C_N/2) f_a^{-1}$; therefore cooling simulations put
constraints on $f_a^*$ rather than on $f_a$ and $C_N$ separately. From
now on, we will also assume that $C_e=0$ -- a conservative assumption
which allows us to focus on PBF processes. We will return to the role
of electrons in a separate study.  All simulations employ the APR EOS
with only nucleonic degrees of freedom, which guarantees that fast
cooling processes do not act. Before presenting the results, we turn
to the observational data.

\subsection{Selecting objects}

As argued previously in KS, it is sufficient to carry out fits to
selected objects rather than a global fit to the population of all
known thermally
emitting neutron stars. Here we use a handful of objects to mark up
the early $\sim 0.1$kyr and intermediate $\sim 100$~kyr evolution of
neutron stars. For the early stages, excellent candidates are the
CCOs  in supernova remnants (SNRs), which
comprise a family of around ten pointlike, thermally emitting x-ray
sources located close to the geometrical centers of nonplerionic
SNRs~\cite{2013ApJ...765...58G}. They do not show counterparts at any
other wavelength than x rays and have low magnetic fields, which
exclude heating processes at this stage of evolution.

As a {\it representative} for CCOs we take the {\it CXO
  J232327.9+584842 } in Cassiopea A SNR. It has received much
attention because of its putative transient cooling claimed to occur
during the past ten years. In the current context these variations are
irrelevant, and we shall adopt a constant temperature
$T = 2.0\pm 0.18 \times 10^6$ K at the age 
$320$~yr~\cite{2013ApJ...777...22E}. As {\it representatives} for late-time
cooling we choose a group of three neutron stars which form a class of
nearby objects that allows spectral fits to their x-ray
emission~\cite{2005ApJ...623.1051D}. Typically the spectra do not
allow a single-component blackbody fit, but two-component fits are
sufficient. The first object is {\it PSR B0656+14}, which is a 
rotation-powered pulsar. The two inferred temperatures for this object are
$T_w = (6.5\pm 0.1)\times 10^5$~K and
$T_h = (1.25\pm 0.03)\times 10^6$~K. The characteristic age of this
pulsar is $1.1 \times 10^{5}$~yr. The second object is {\it PSR
  B1055-52} which is again a rotation powered
pulsar~\cite{2005ApJ...623.1051D}. The two black-body temperature fits
give $T_w = 7.9\pm 0.3\times 10^5$~K and
$T_h = (1.79\pm 0.06) \times 10^6$~K. The characteristic age of this
pulsar is $5.37 \times 10^{5}$~yr. The third object is {\it Geminga},
which is a radio-quite nearby x-ray-emitting neutron
star~\cite{2005ApJ...623.1051D}. The two-blackbody temperature fit
gives $T_w = 5.0\pm 0.1\times 10^5 $~K and
$T_h = (1.9\pm 0.3) \times 10^6$~K.  The characteristic age of Geminga
is $3.4 \times 10^{5}$ yr.  In confronting the neutron stars'
blackbody temperatures with the theoretical models, we will adopt the
lower of the two values inferred.  The ages of these three neutron
stars are known only on the basis of a spin-down model, which is
uncertain. We quantify this uncertainty by assigning a factor of 3
error to the spin-down age of each of these objects.  The data on PSR
B1055-52 are marginally (in)consistent with the cooling curves we
find, but the uncertainties in the physics of cooling tolerate such
discrepancy: first, in contrast to CCO in Cas A, only the spin-down
age is known, which can have larger error than we assumed; second,
at the later stages of thermal evolution, heating processes (even for
weakly magnetized stars) can become a factor. Finally, the discrepancy
may lie in the modeling of the pairing gaps, which can be tuned to fit
the inferred temperature of PSR B1055-52.

\subsection{Results of simulations}

The results of extensive simulations are summarized in
Fig.~\ref{fig:2}, where we show cooling tracks for 16 models of
 light- and intermediate-   ($m=1$ and $m=1.4$) mass neutron stars for
 cases of  a nonaccreted iron envelope  ($\eta = 0$) and a light-element
 envelope  ($\eta = 1$). For each of these cases the axion
coupling has been assigned the following values: ~$f_a = \infty$
(negligible  coupling), $f_{a7} = 10$, $f_{a7} = 5$ and   $f_{a7} =
2$, where $f_{a7} = f_a/10^7$ GeV, in combination with charges
$\vert C_n\vert= 0.04$  and  $\vert C_p\vert = 0.4$.

The observational temperatures of the four objects discussed are shown
by dots with error bars. The temperature of CCO in Cas A is consistent
with the cooling of $m=1$ and 1.4 mass stars assuming that the compact
object in Cas A has a light-element envelope and axion cooling is
absent. Switching on the axion cooling decreases the temperatures of
models with the age of CCO in Cas A because of the additional losses
caused by the axion PBF process. It is seen that for small enough
values of $f_a$, the cooling curves become inconsistent with the Cas A
data.  Quantitatively, the lowest value of axion coupling $f_{a7} = 2$
is inconsistent with both $m=1$ and $m=1.4$ mass cooling; the value
$f_{a7} = 5$ is inconsistent with $m=1$ but not with $m=1.4$ mass star
cooling.
\begin{figure}[tbh] 
\begin{center}
\includegraphics[width=8cm,height=7cm]{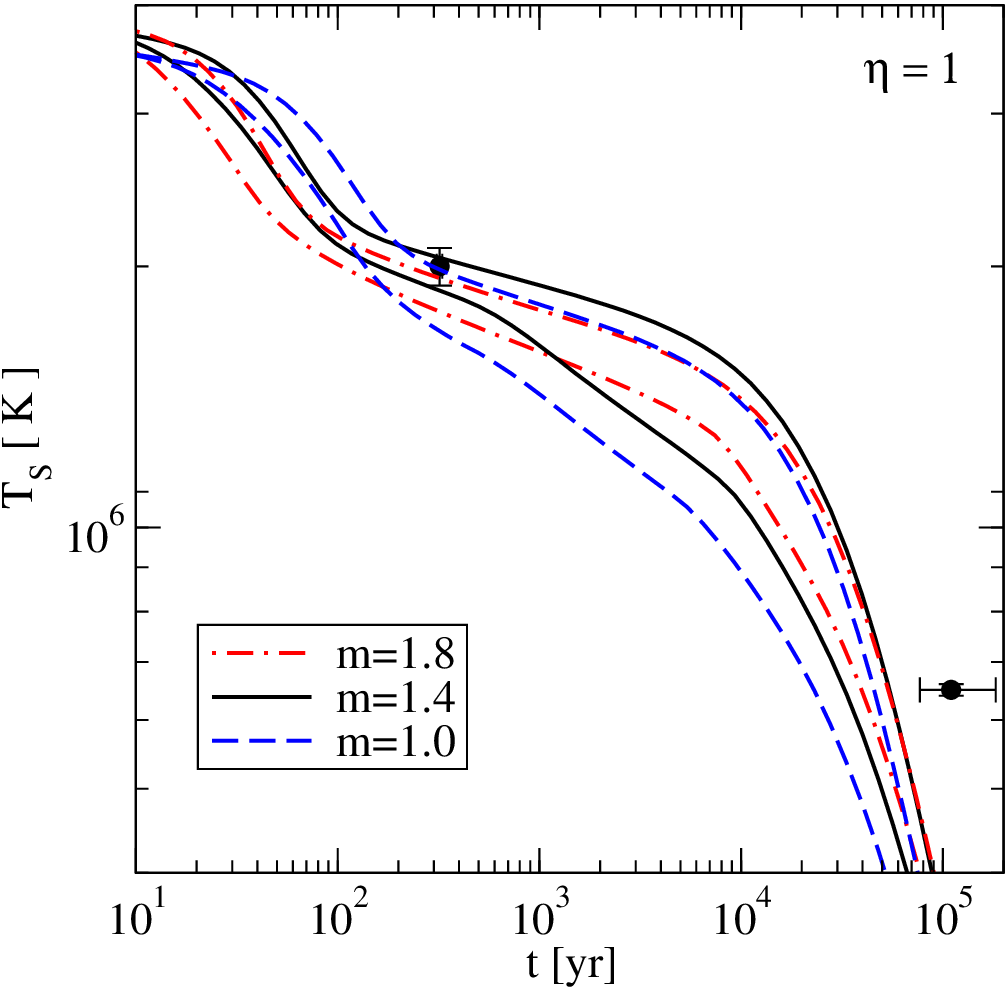}
\caption{   Cooling tracks of neutron star models with 
  masses $m=1$ (dash-dotted) 1.4 (solid) and 1.8 (dashed) for the case 
  of an accreted light-element envelope ($\eta=1$) along with the 
  measured temperature of CCO in Cas A. For each value of mass, the 
  upper curve corresponds to the cooling without axions, and the lower
  curve corresponds
  to  axion cooling with $f_{a7} = 5$. Note the weak dependence on the 
  surface temperature of models on the star mass.  }
\label{fig:3}
\end{center}
\end{figure}
\begin{figure}[!] 
\begin{center}
\vskip 0.5cm 
\includegraphics[width=8cm,height=7cm]{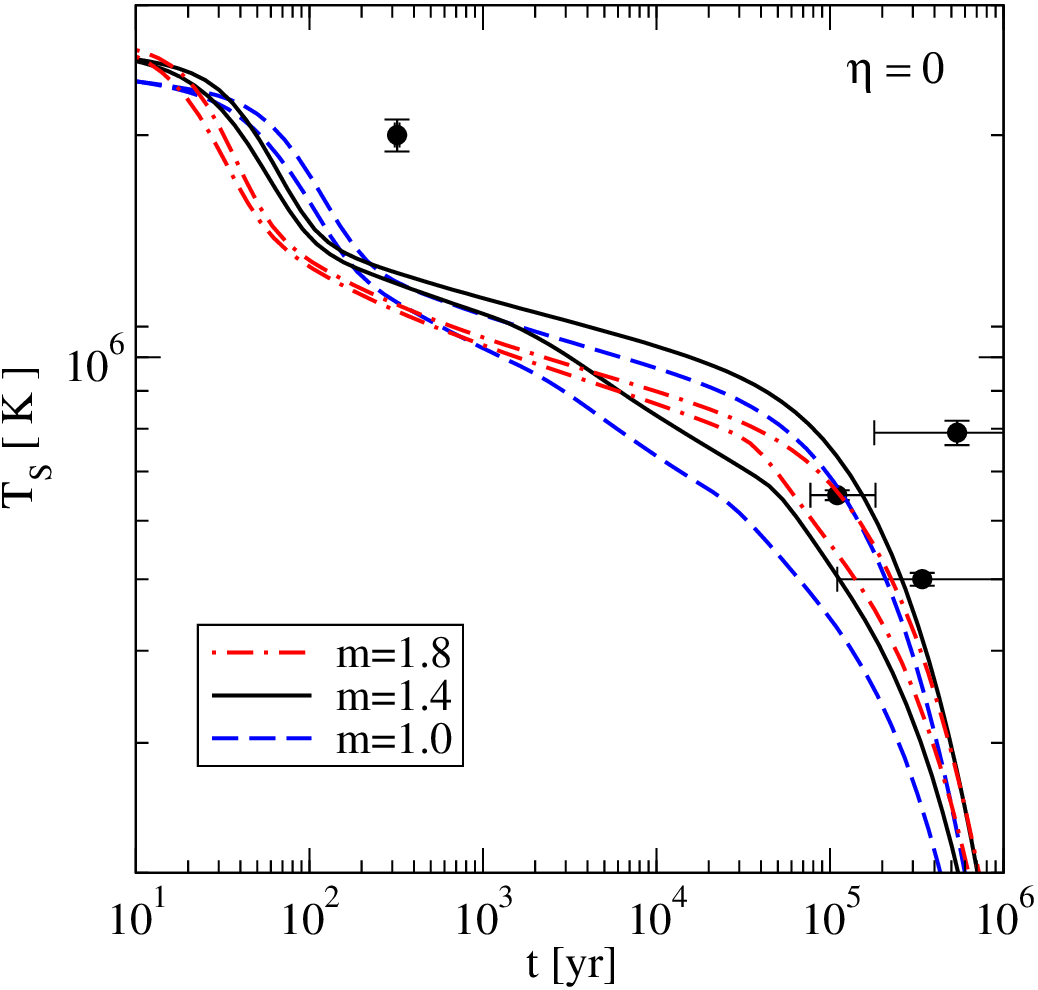}
\caption{   Cooling tracks of neutron star models with
  masses $m=1$ (dashed) 1.4 (solid), and 1.8 (dash-dotted) for the
  case of a nonaccreted iron envelope ($\eta=0$). The measured
  temperatures of PSR B0656+14, Geminga are consistent with neutrino
  cooling tracks; the uncertainty in the spin-down age of PSR B1055-52
  and internal heating may account for marginal inconsistency. The
  axion cooling tracks are shown for $f_{a7} = 10$. }
\label{fig:4}
\end{center}
x\end{figure}
\begin{figure}[tbh] 
\begin{center}
\includegraphics[width=8cm,height=7cm]{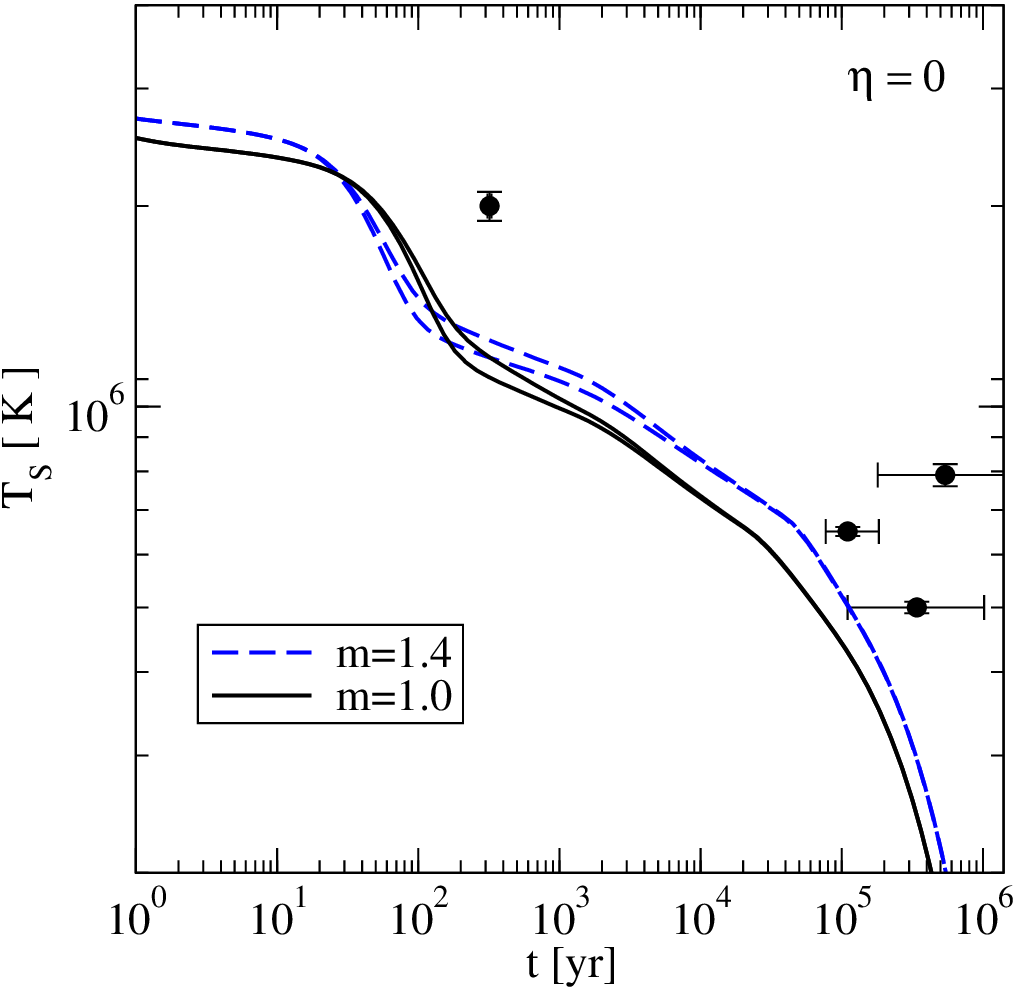}
\caption{  Cooling tracks of neutron star models  with 
  masses $m=1$ (solid) and 1.4 (dashed) for the case of a nonaccreted
  iron envelope ($\eta=0$) and for $f_{a7} = 10$. For each mass, the
  two tracks differ by the value of the neutron PQ charge.  The upper
  curves correspond to our standard choice $\vert C_n \vert= 0.04$,
  while the lower curves correspond to the case of enhanced axion
  emission with $\vert C_n \vert= \vert C_p \vert= 0.4$.}
\label{fig:5}
\end{center}
\end{figure}
\begin{figure}[tbh] 
\begin{center}
\includegraphics[width=8cm,height=7cm]{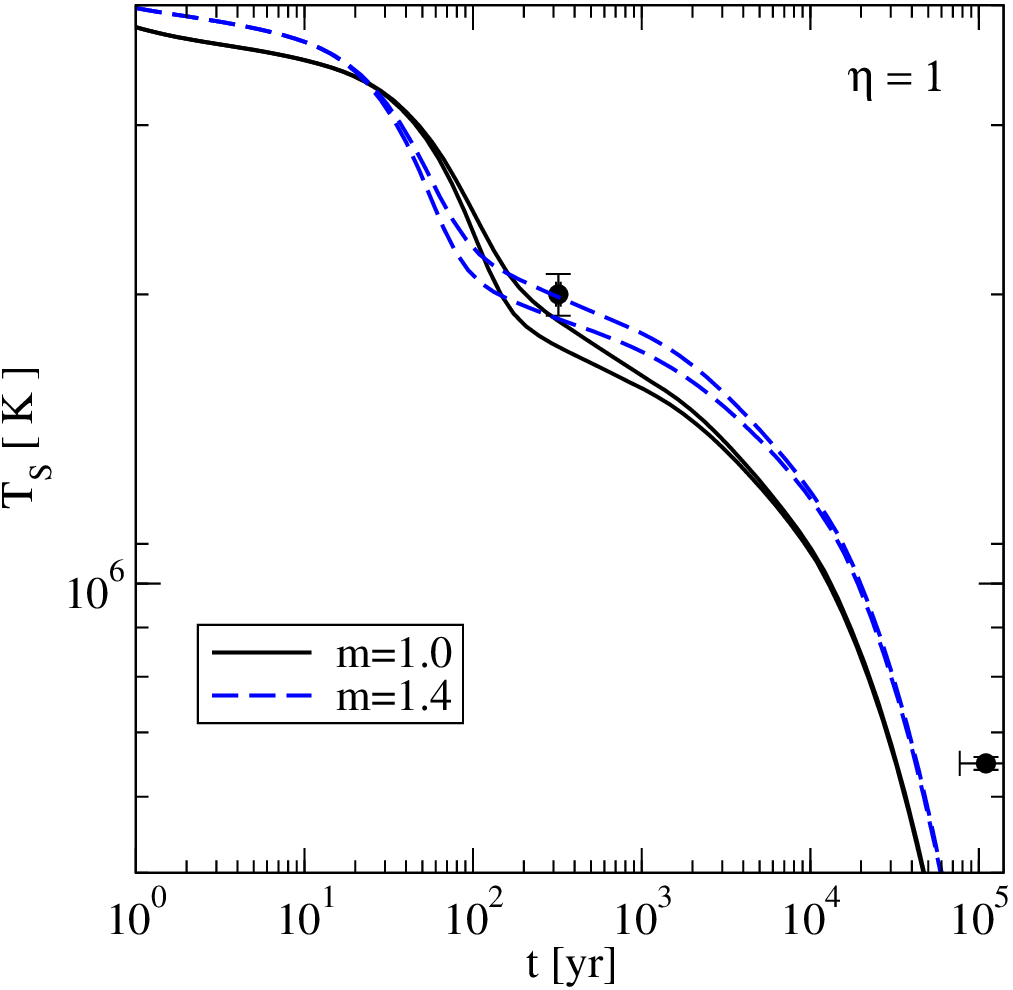}
\caption{  Same as in Fig.~\ref{fig:5}, but in the case
of an accreted envelope ($\eta=1$) .
}
\label{fig:6}
\end{center}
\end{figure}

The temperatures of the remaining middle-aged neutron stars from our
collection are consistent with  the cooling of $m=1$ and 1.4 mass star
models if we make the natural assumption that these neutron stars have
nonaccreted iron envelopes. Axion cooling with $f_{a7} \le 5$ is
clearly inconsistent with the data on these objects. For $f_{a7} = 10$
and $m=1.4$, the cooling tracks are marginally consistent with the
data. Physically, the inconsistency arises from the PBF axion cooling
of the models {\it prior} to the actual age of these objects, which
according to simulations are currently cooling predominantly 
via crust bremsstrahlung and surface photon emission.

Figure~\ref{fig:3} focuses on the cooling behaviour at the early
stages of evolution and on CCO in Cas A. Here we have added also cooling
tracks for massive $m=1.8$ stars to quantify the variations in the
mass of the objects. It is seen that significant variations in the
mass do not change the cooling tracks; this would, of course, change
if the EoS of dense matter admits fast cooling processes -- i.e., if in
more massive stars the threshold densities for the onset of rapid
cooling processes are attained.  Our computations show that
$f_{a7} = 10$ cooling is still consistent with the data for $m=1.4$
stars, but for $f_{a7} = 5$ cooling tracks are inconsistent with the
data independent of the mass of the star, as shown in
Fig.~\ref{fig:3}.

Figure~\ref{fig:4} focuses on the cooling of the three 
intermediate-aged 
neutron stars discussed above with and without axion cooling. The
variation in the mass range $1\le m\le 1.8$ does not induce significant
changes in the cooling tracks, provided that fast cooling processes do
not operate in the massive $m=1.8$ model.  The data are
consistent with neutrino-only cooling,
assuming that some minor adjustment can improve
the agreement with PSR B1055-52 data. (We recall that the spin-down
age may have larger error than assumed, or some heating processes may
already operate in this object.)  Turning on the axion cooling it is
seen that $f_{a7} = 10$ cooling tracks are clearly inconsistent with
the data, independent of the mass of the star.  

To conclude, the combination of observational data and simulations
including cases with nonaccreted ($\eta=0$) and accreted ($\eta =1$)
envelopes suggests that the range of axion coupling constant for which
axion cooling is inconsistent with data lies within
$5\le f_{a7} \le 10$ independent of the mass of the star.

So far, we have fixed the values of PQ charges of the neutron and proton to
some characteristic values taken from the range defined by the
inequalities \eqref{eq:axion_range}.  The proton PQ charge is
constrained in this class of theories to a narrow range of values,
while the neutron PQ charge changes the sign, thus allowing for zero
coupling of the axion to the neutron. Furthermore, the PQ charge of
neutron is at least a factor of 4 smaller than the proton charge. For the
value of $\vert C_n\vert =0.04$ we adopted and for our choice of
pairing gaps, the axion emission is dominated by proton condensate, and
the emission from neutron condensates is negligible.  To see the
possible effect of the neutron condensate on the cooling evolution, in
particular on the range of the temperatures and the time span where it
may play a role, we have simulated the cooling with a model where
$\vert C_n\vert =\vert C_p\vert =0.4$.
Figure \ref{fig:5} shows the cooling models in the case of
nonaccreted envelops ($\eta =0$) for light neutron star models and
the value of axion coupling $f_{a7} = 10$.  For each mass we show two
cooling curves with $ \vert C_n \vert =0.04$ and
$\vert C_n \vert =0.4$ (where the upper curve always corresponds to
the small value of $\vert C_n \vert$).  The same, but in the case of
accreted envelopes ($\eta =1$), is shown in Fig.~\ref{fig:6}. Axion emission
by neutron condensate lowers the surface temperatures of the models by
a factor of the order of unity; therefore, the cases where
$\vert C_n\vert \sim \vert C_p\vert $ will have qualitatively similar
bounds to those obtained above. It is seen that that the neutron
condensate affects cooling during the time span $10^2\le t\le 10^3$
yr, which corresponds to interior temperatures in the range
$0.5 \, T_c\le T < T_c$. The role of the neutron condensate can become
important if fine-tuning of the cooling curves to  data will be
required, as is possibly the case for the CCO in Cas A (see
Ref.~\cite{2015arXiv150906986S} and references therein).

\section{Discussion and conclusions}
\label{sec:conclusions}

This work explores how the emission of axions by weakly magnetized
neutron stars during their early ($t\sim 0.1$ kyr) and intermediate
($t\sim 10^2$ kyr) evolution alters their observable surface
temperatures. As a benchmark, we modeled the purely neutrino cooling of
neutron stars within a slow cooling scenario where any fast cooling
processes, such as the direct Urca processes on nucleons and quarks,
are excluded. These purely neutrino cooling models are consistent with
the temperature of CCO in Cas A if we assume this object has a
light-element envelope; these cooling tracks are also consistent with
the older pulsars and Geminga if we assume a nonaccreted, iron envelope
and account for errors in the age determinations and possible changes
due to internal heating. The dependence of the
cooling tracks on the mass of the models is rather weak because of
absence of fast cooling agents. We further explored the influence of
axion cooling bremsstrahlung processes on the cooling tracks of our
models by smoothly varying the axion coupling constant $f_a$ (the
strength of the coupling scales as $1/f_a$). In doing so, we fixed the
PQ charges of the neutron and proton at the values $\vert C_n\vert = 0.04$ 
and $\vert C_p\vert = 0.4$ motivated by hadronic models of axions [see
Eq. \eqref{eq:axion_range}] and
neglected the coupling of the axion to electrons, $C_e = 0$ (this
would correspond to the KVSZ class of models of axions). The latter
conservative assumption strengthens the limits, because the inclusion
of axion emission by electron bremsstrahlung processes would have
increased the discrepancy between the models and purely
neutrino cooling models. We find that the value of  $f_{a7} = 5$
is clearly inconsistent with the combined observational data, and 
 $f_{a7} = 10$ is inconsistent with the surface temperatures of
 middle-aged neutron stars.  Using these bounds in the 
relation \eqref{eq:axion_mass}, we obtain the following conservative
limit on the axion mass:
\bea 
\label{eq:axion_mass2}
f_a/10^{7}\textrm{GeV} \ge (5\textrm{--}10),\quad 
m_a \le (0.06\textrm{--}0.12)~\textrm{eV}. 
\eea
which can be contrasted with the bound given by KS \eqref{eq:mbound}
for the value $C_N = \vert C_p\vert  = 0.4$:
\bea 
\label{eq:axion_mass3}
f_a/10^{7}\textrm{GeV} \ge 15.2,\quad m_a \le 0.04~\textrm{eV}. 
\eea

The obtained upper bound on the mass of the axion is consistent with
those obtained from the
supernova~\cite{1988PhRvD..38.2338B,1989PhRvD..39.1020B,1990PhRvD..42.3297B,1996PhRvL..76.2621J,2001PhLB..499....9H}
and proton-neutron star \cite{1990PhRvD..42.3297B} physics,
$m_a \le 0.1 $eV.  However, the limit
\eqref{eq:axion_mass2} is  based on a rather conservative segment of
the physics of cooling of neutron stars and surface temperature data
measured from nearby x-ray-emitting neutron stars and is complementary
to the one quoted above. The bounds derived from proto–neutron stars
share the same type of uncertainties as the cold neutron star model,
while the supernova limits suffer from the uncertainty in the basic
mechanism that drives supernova explosions.  Limits similar to ours
were obtained by Umeda {\it et al.}~\cite{1998nspt.conf..213U} in their
pioneering study of the axion cooling of neutron stars, although their
study does not include the key PBF processes; i.e., their axion
cooling is dominated by nucleon bremsstrahlung processes.

Looking ahead, it should be mentioned that the present study selected
only a single pair of values of PQ charges for neutrons and protons from
a range defined for hadronic models of axions. In general these
charges may vary and thus define a continuum of axion models. A broader
overview of the axion cooling of neutrons can be obtained by varying
independently these two parameters, as well as by fixing their values
to specific models such as the DVSZ or the KVSZ models. A further point
for a future study is the role of the axion bremsstrahlung by
electrons. Electron bremsstrahlung of axions in the crust at later
stages of neutron star cooling (which we neglected in this study
assuming $C_e=0$, as in the KVSZ model) needs to be included in the
theoretical models of axion cooling. This will allow us to improve the
constraints on the axion mass, as additional axion emission will lead to
discrepancies between the theory and observations at larger values of
$f_a$ than quoted above.

\section*{Acknowledgments}

The support of this research by the Deutsche Forschungsgemeinschaft
(Grant No.~SE 1836/3-1) and by NewCompStar COST Action MP1304 is
gratefully acknowledged.


\end{document}